\documentclass[%
 aip,
 jcp,
 amsmath,amssymb,
 reprint,%
floatfix
]{revtex4-2}

\usepackage{graphicx}
\usepackage{dcolumn}
\usepackage{bm}
\usepackage[T1]{fontenc}
\usepackage{mathptmx}
\usepackage{etoolbox}
\usepackage[inline]{enumitem}
\usepackage{chemformula}
\usepackage{xcolor}

\usepackage{amssymb}

\makeatother

\begin{document}

\title{Signatures of auxeticity in microgels at low and ultralow crosslinker concentration}


\author{Susana Mar\'in-Aguilar}
\email{susana.marinaguilar@uniroma1.it}
\affiliation{Department of Physics, Sapienza University of Rome, Piazzale Aldo Moro 2, 00185 Roma, Italy.}

\author{Leah Rank}
\affiliation{Department of Physics, Sapienza University of Rome, Piazzale Aldo Moro 2, 00185 Roma, Italy.}
\affiliation{CNR Institute of Complex Systems, Uos Sapienza, Piazzale Aldo Moro 2, 00185, Roma, Italy.}

\author{Emanuela Zaccarelli}
\email{emanuela.zaccarelli@cnr.it}
\affiliation{CNR Institute of Complex Systems, Uos Sapienza, Piazzale Aldo Moro 2, 00185, Roma, Italy.}
\affiliation{Department of Physics, Sapienza University of Rome, Piazzale Aldo Moro 2, 00185 Roma, Italy.}
 
\date{\today}

\begin{abstract}
Auxetic behavior, characterized by a negative Poisson's ratio, is a counterintuitive mechanical response exhibited, among other systems, by certain polymer networks. Here, through \textit{in silico simulations}  we investigate the mechanical response of thermoresponsive microgels across the volume phase transition
upon varying crosslinker concentration down to ultralow conditions, a regime so far unexplored.
After refining the method to estimate the elastic moduli based on equilibrium shape fluctuations for the challenging case of ULCs, which are very sparse networks with rather anisotropic shape, we are able to show the onset of auxetic behavior near the volume phase transition for microgels with crosslinker concentration of $\sim 1\%$. In addition, we find that ULC microgels exhibit a slightly negative Poisson's ratio across the whole swollen regime. Further examining the auxetic response within the inner region of the network, we also demonstrate that, for ULC microgels, this extends at all length scales, suggesting that it is an intrinsic property of the weakly connected polymer network. The present findings should likely stimulate novel experimental investigations, aiming to measure the Poisson's ratio of individual low and ultralow crosslinked microgels, to verify these intriguing numerical predictions.
\end{abstract}

\maketitle

\section{Introduction}

Auxeticity defines the mechanical behavior for which a material exhibits a negative Poisson's ratio $\nu$. Contrary to standard expectations, when stretched (compressed), it expands (contracts) in the transversal direction. This behavior has been reported in a wide range of systems, including foams, polymers, crystalline solids, and biological materials~\cite{Lakes1987, Caddock1989, Evans1989, Evans1991, Evans2000, Greaves2011, Gatt2015, Zulifqar2016, Rysaeva2019}, paving the way for various applications in medicine and protective materials~\cite{Bose2012, Duncan2018, Tahir2022}. From a mechanical standpoint, auxeticity is typically associated with the relative softening of the bulk modulus $K$ as compared to the shear modulus $G$, leading to enhanced volume fluctuations under deformation and hence to a negative value of the Poisson's ratio $\nu$~\cite{Landau1960Theory}.

Among the systems that exhibit auxetic behavior, we find thermoresponsive polymer networks based on poly(N-isopropylacrylamide) (pNIPAM). In particular, pioneering studies reported the onset of auxetic behavior in pNIPAM hydrogels~\cite{Hirotsu1991, Li1993, Hirotsu1994} close to the so-called volume phase transition (VPT). This transition, which manifests as a sudden collapse of the network, is driven by the underlying coil-to-globule transformation of polymer chains~\cite{Lifshitz1969} due to
the worsening of polymer-solvent affinity at high temperatures. In addition to bulk systems, finite-sized colloidal networks, such as microgels, have also been found to exhibit a change in their mechanical properties close to the VPT. In particular, capillary micromechanics ~\cite{voudouris2013micromechanics} and atomic force microscopy (AFM) measurements~\cite{hashmi2009mechanical,burmistrova2011effect} have reported the presence of a minimum in the bulk or Young's modulus for individual microgels. This is accompanied by the presence of a minimum in the Poisson's ratio behavior with temperature, occurring close to the VPT temperature~\cite{voudouris2013micromechanics}. This finding was later confirmed by numerical studies of elastic properties of single microgels~\cite{rovigatti2019connecting}, which also showed that 
$\nu$ decreases with decreasing crosslinker concentration $c$. 

In all the previous microgel studies, the Poisson's ratio was found to be positive. Instead, a negative value of $\nu$ was reported in simulations of hydrogels for low values of $c$~\cite{Ninarello2022} at small negative pressures. This corresponds to the application of a tension, which stretches the chains composing the network and thus counteracts the low connectivity of the network, until a mechanical instability is detected for ultra-low-crosslinked conditions for which $\nu = -1$, i.e., the lowest attainable value in 3D~\cite{Greaves2011}. Such instability was found to occur both for ordered (diamond-like) and for disordered networks~\cite{Ninarello2023}, suggesting that the underlying topology 
does not play an important role. Rather, 
the key control parameter governing auxeticity seems to be related to network connectivity. These findings agree with the general understanding that auxetic behavior can emerge not only through geometric design~\cite{grima2005auxetic, Bertoldi2017, Larsen1997, Theocaris1997, Hanifpour2017, Reid2018}, but also from thermodynamic effects~\cite{Hirotsu1994,Hirotsu1991,Li1993,dong2010softening,lakes2017negative,Ninarello2025}.

To recap, so far, auxetic behavior in polymeric hydrogels was previously found at very low $c$ upon applying a tension in simulations~\cite{Ninarello2022} and theory~\cite{grima2000auxetic} or in experiments upon varying temperature~\cite{Hirotsu1991}. Conversely, in microgels, which by definition are in equilibrium at zero osmotic pressure, $\nu$ was found to decrease with $c$, never reaching negative values. Merging these two behaviors conceptually, one would think that the Poisson's ratio of very-low-crosslinked microgels may become negative close to the VPT at zero pressure. This should be even more true for ultra-low-crosslinked pNIPAM microgels, which are synthesized in the absence of any crosslinkers and stabilized by rare self-binding events among NIPAM chains~\cite{gao2003cross}.

To this aim, in this work, we investigate the elastic properties of individual microgels with different crosslinker concentrations at zero external pressure. We rely on monomer-resolved simulations~\cite{gnan2017silico,ninarello2019modeling}, able to describe the internal structure of experimental systems down to ultra-low-crosslinking~\cite{hazra2023structure}. First of all, we adapt the method, based on volume and shape fluctuations~\cite{rovigatti2019connecting}, to extract the elastic moduli to the case of sparsely-connected microgels with long dangling chains. After validating it, we provide numerical estimates of bulk and shear moduli of the microgels as a function of temperature. From these, we then extract the Poisson’s ratio and find that this exhibits a minimum at the VPT, reaching auxetic conditions at low enough crosslinker concentrations. Moreover, for ULC microgels, the auxetic behavior is found to persist at all temperatures below the VPT.
Furthermore, we examine the spatial distribution of the elastic response inside the microgel, finding that auxeticity in ULC microgels is an intrinsic property that belongs to the whole particle, while it is limited to the outer corona at finite $c$.
Altogether, these findings suggest that, for low-crosslinker microgels, auxeticity emerges from the interplay between thermal swelling and network sparsity, awaiting experimental confirmation in the near future.

\section{Models and Methods}

\subsection{Molecular Dynamics Simulations}
To investigate the elastic properties of microgels with various crosslinker concentrations, $\%c$, we rely on the {\it in silico} synthesis of microgels introduced in Refs.~\cite{gnan2017silico, ninarello2019modeling}. First, we assemble a microgel with $N$ monomers, including a fixed number of crosslinkers $N_c$. The total number of particles is $N\sim42000$ monomers. To assemble the microgels, we perform  NVT molecular dynamics (MD) simulations of a binary mixture of patchy particles confined within a sphere of radius $Z$. Particles with two and four patches represent monomers and crosslinkers, respectively. The assembly simulations are carried out using the OxDNA package~\cite{poppleton2023oxdna}. Additionally, at this stage, a radial force is applied to the crosslinkers, following Ref.~\cite{ninarello2019modeling}, in order to mimic the microgel's heterogeneous internal structure. 
In this way, we assemble microgels with $\%c=1.0$ and $5.0$, by fixing $Z$ so that the microgel has a number density of $\rho_\mathrm{st}\sim0.08$. In addition, we also investigate ultra-low-crosslinked microgels (ULCs)~\cite{hazra2023structure, marin2026unexpected}. For these systems, the assembly is performed using a binary mixture of patchy particles with three (crosslinkers) and two patches, using no addition of radial force. This choice is made to reproduce the low connectivity characteristic of ULC microgels. The ULC crosslinker concentration is fixed to $c=0.1\%$, while the single microgel density is reduced to $\rho_\mathrm{ULC}\sim0.029$, to reproduce experimental form factors, as reported in Ref.~\cite{hazra2023structure}. 

Once the network is almost fully connected, with around $\sim99\%$ of particles bonded within one big network, the patchy interactions are replaced by bead-spring bonds. This fixes the structure of the microgels, mimicking the effect of covalent bonds. Consequently, all particles interact through the Weeks-Chandler-Andersen (WCA) potential, 

\begin{equation}
    U_{\mathrm{WCA}}(r)=\begin{cases} 4\epsilon\left[ \left(\frac{\sigma}{r}\right)^{12}-\left(\frac{\sigma}{r}\right)^6 \right]+\epsilon & r\leq 2^{1/6}\sigma \\
    0 & r> 2^{1/6}\sigma,
    \end{cases}
\end{equation}
with $r$ being the distance between two monomers and $\sigma$ the diameter of each bead; the latter is used as the unit of length, and $\epsilon$, the strength of the interaction, sets the unit of energy. The bonded monomers, interact through the Finite-Extensible-Nonlinear-Elastic (FENE) potential~\cite{kremer1990dynamics}:
\begin{equation}
    U_{\mathrm{FENE}}(r)=-\epsilon k_F {R_0}^2\log{\left[ 1-\left(\frac{r}{R_0\sigma}\right)^2 \right]} \text{ if } r<R_0\sigma,
\end{equation}
with $R_0=1.5$ and $k_F=15$, setting the maximum bond extension and stiffness, respectively. 

We are interested in exploring the behavior of the elastic response across the Volume Phase Transition (VPT). Therefore, we include an effective attractive interaction between the monomers, which represents the gradual change from solvophilic to solvophobic interactions experienced by the pNIPAM monomers upon increasing temperature~\cite{soddemann2001generic}. This potential has the following form,
\begin{equation}
    U_{\alpha}(r)=\begin{cases}
    -\epsilon \alpha & r\leq 2^{1/6}\sigma \\
    \frac{1}{2}\alpha\epsilon\left[\cos{\left(\gamma\left(\frac{r}{\sigma}\right)^2+\beta\right)-1}\right] & 2^{1/6}\sigma<r\leq R_0, \\
    0 & r> R_0
    \end{cases}
    \label{eq:alpha}
\end{equation}
where $\gamma=(\pi(2.25-2^{1/3}))^{-1}$, $\beta=2\pi-2.25\gamma$, and $\alpha$ tunes the strength of the attractive interaction, thus amounting to an effective temperature: $\alpha=0$ corresponds to good solvent conditions, while $\alpha\sim0.63$ roughly coincides with the VPT temperature~\cite{Ninarello2025}. For $\alpha> 1$ the microgel is fully collapsed.

We perform MD simulations of single microgels using LAMMPS~\cite{thompson2022lammps} with a time step of $\delta t=0.002\tau$, where $\tau=\sqrt{m\sigma^2/\epsilon}$ corresponds to the MD time unit. For each $\%c$ and $\alpha$, the system is equilibrated for at least $t=5\times10^6\tau$, followed by a production run ranging from $t=1\times10^7-5\times10^7 \tau$. During the production stage, configurations are sampled every $5000$ time steps to accurately determine the elastic moduli. Finally, to improve statistics and to account for different topologies, we consider at least three independent configurations per crosslinker concentration. For ULC microgels, instead, we simulate $10$ different configurations, due to their large structural fluctuations.

To determine the different microgel sizes, we calculate their hydrodynamic radius, $R_h$, as
\begin{equation}
    R_h = \Bigg\langle2\Bigg[ \int_{0}^{\infty}\frac{1}{\sqrt{(a_1^2 + \theta)(a_2^2 + \theta)(a_3^2 + \theta)}} \, d\theta \Bigg]^{-1}\Bigg\rangle,
    \label{eq:rh}
\end{equation}
with $a_i$ the principal semiaxes of the gyration tensor associated with a convex hull enclosing the microgel (see the following Section for extensive details). This method was put forward by Hubbard et al.~\cite{Hubbard1993} to obtain the hydrodynamic radius of arbitrarily shaped Brownian particles and was later validated for regular microgels against experiments in Ref.~\cite{del2021two}.

To characterize the internal structure of the various microgels, we calculate the radial density profiles, $\rho(r)$, as
\begin{equation}
    \rho(r)=\frac{1}{V}\left<\frac{1}{N}\sum_{i=1}^N \delta(|\mathbf{r}_i-\mathbf{r}_{cm}|-r)\right>,
    \label{eq:densprof}
\end{equation}
where $\mathbf{r}_{cm}$ corresponds to the position of the center of mass of the microgel and $\mathbf{r}_i$ is the position of monomer $i$. The resulting profiles are fitted using the so-called fuzzy-sphere model,

\begin{equation}
    \rho(r)\sim A  \textrm{ erfc}\left(\frac{r-R_c}{\sqrt{2}\sigma_s}\right),
    \label{eq:fuzzysphere}
\end{equation}
where $A$ is a fit parameter, while $R_c$ and $\sigma_s$ correspond to the core radius and the half-width of the corona, respectively. Although ULC microgels possess a much looser internal structure than standard microgels and lack a defined core, their radial density profiles are still well described by the fuzzy-sphere model~\cite{marin2026unexpected}.

\subsection{Elastic Moduli}

To determine the elastic moduli of the microgels, we follow the approach established in Ref.~\cite{rovigatti2019connecting} based on Ref.~\cite{aggarwal2016nonuniform} and on the Mooney-Rivlin theory~\cite{treloar1976mechanics}. This method relates the structural fluctuations of a single microgel over time to the corresponding elastic response through isotropic strain invariants. To calculate such invariants, we build a convex hull that encloses all monomers around each sampled equilibrated microgel configuration. The gyration tensor of the convex hull surface is then calculated from the center of mass of each triangle defining the convex hull. With this, we obtain the corresponding eigenvalues $\lambda_i$, which can be related to the three semi-axes of an ellipsoid through $a_i=\sqrt{3\lambda_i}$. 

We also tested an alternative representation of the microgel surface based on the $\alpha$-shape method implemented in OVITO~\cite{Stukowski2010, Stukowski2014}. This method generates a triangulated representation of the microgel surface from a Delaunay tessellation of the particle coordinates. The resulting surface is defined through an $\alpha$-shape construction employing a probe sphere of radius $R_{\mathrm{\alpha}}$. The value of $R_{\mathrm{\alpha}}$ controls the level of detail retained in the surface mesh. Smaller radii capture finer geometric features, whereas larger radii produce a smoother and more convex envelope. In the limit of sufficiently large $R_{\mathrm{\alpha}}$, the reconstructed shape approaches the convex hull. Analogously to the description above, the gyration tensor is built using now the points obtained from the surface mesh, and the corresponding three semi-axes are retrieved. Although the ($\alpha$)-shape method provides a more detailed representation of the microgel boundary, we found that the associated shape fluctuations exhibit significantly larger statistical noise, leading to less reliable estimates of the elastic moduli, as shown below. 

With the use of the related ellipsoidal semi-axes $a_i$, either obtained from the convex-hull or the surface mesh, we then build the Green-Lagrange strain tensor $\mathbf{C}=\mathbf{F}^T\cdot\mathbf{F}$, where $\mathbf{F}$ is a diagonal matrix built from the fluctuations around a reference configuration. The latter is assumed to correspond to the average of the semi-axes $\left< a_i\right>$. Therefore $\mathbf{F}$ reads, 
\begin{equation}
    \mathbf{F}=
    \renewcommand\arraystretch{1.5}
    \begin{pmatrix} 
    \frac{a_1}{\left< a_1\right>} & 0 & 0\\
    0 & \frac{a_2}{\left< a_2\right>} & 0 \\
    0 & 0 & \frac{a_3}{\left< a_3\right>}\\
    
    \end{pmatrix}
\end{equation}

From $\mathbf{C}$, three rotationally strain invariants are calculated~\cite{aggarwal2016nonuniform},

\begin{equation}
    J=\sqrt{\det(\mathbf{C})}
\end{equation}

\begin{equation}
    I_1=\mathrm{tr}(\mathbf{C})J^{-2/3}
\end{equation}

\begin{equation}
     I_2=\frac{1}{2}(\mathrm{tr}^2-\mathrm{tr}(\mathbf{C}^2))J^{-4/3}.
\end{equation}
The Green-Lagrange strain tensor, $\mathbf{C}$, is equal to the identity matrix in the reference configuration, yielding $J=1$ and $I_1=I_2=3$. 
The total elastic strain energy is expressed in terms of these invariants as~\cite{aggarwal2016nonuniform}, 
\begin{equation}
    U(\mathbf{C})=\bar{U}_0+\int_VW(I_1,I_2,J) dV, 
\end{equation}
where $\bar{U}_0$ is an arbitrary reference energy and $W$ the strain-energy function. Following the Mooney-Rivlin theory on rubber elasticity~\cite{treloar1976mechanics}, $U$ can be expressed as,
\begin{equation}
\begin{array}{lcl}
    U&=&\bar{U}_0+W(I_2)+W(I_1)+W(J)\\ 
    &=&\bar{U}_0+V\left[C_{01}(I_2-3)+C_{10}(I_1-3)+D_1\frac{(J-1)^2}{J}\right], 
\end{array}
\label{eq:invariants}
\end{equation}
with $C_{01}$, $C_{10}$ and $D_1$ corresponding to elastic constants, and $V$ the volume of the reference configuration obtained from the volume of the average ellipsoid, $V=\frac{4}{3}\left<a_1\right>\left<a_2\right>\left<a_3\right>$. The elastic constants can be related to the bulk modulus as $K=2D_1$, and the shear modulus $G=2(C_{10}+C_{01})$. 

\begin{figure*}[th!]
\begin{center}
\includegraphics[width=1\linewidth]{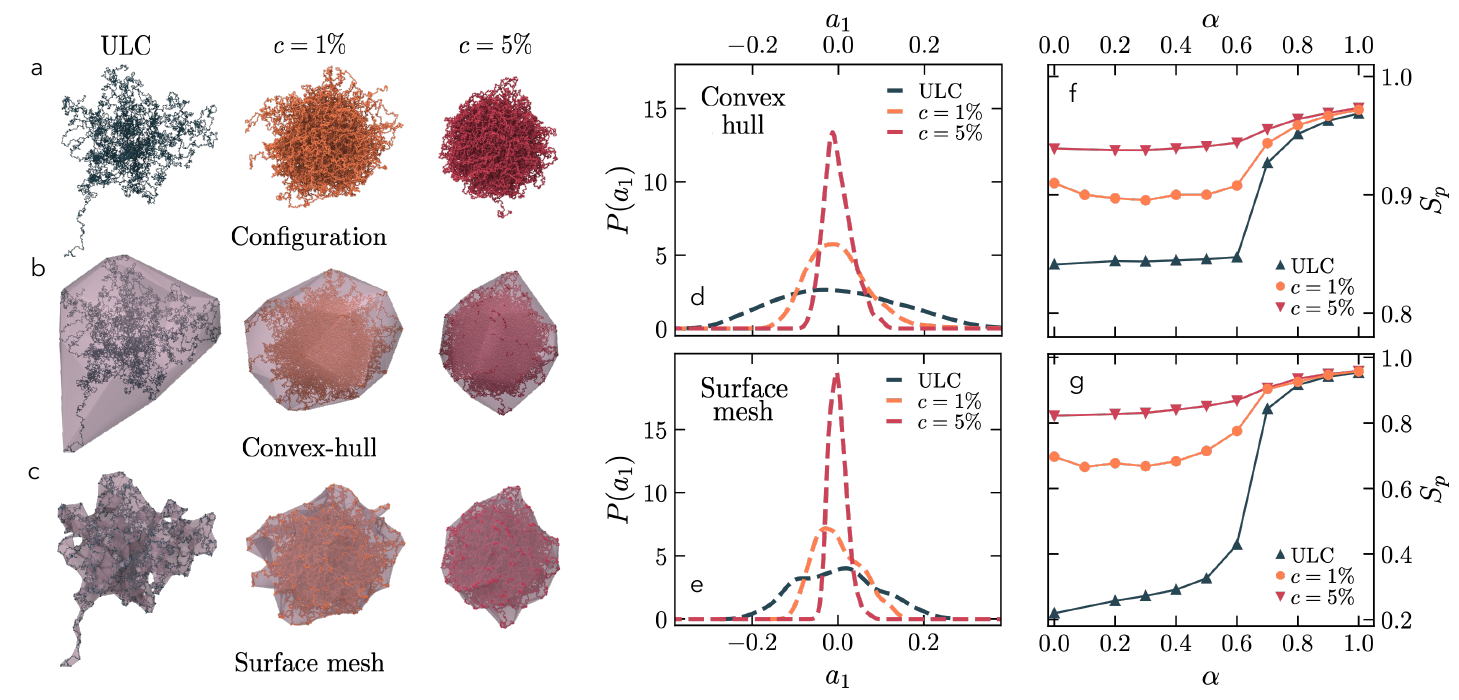}
\end{center}
\vspace{-0.5cm}
\caption{
\textbf{Differences from shape characterizations} a) Representative swollen configurations at $\alpha=0$ of ultra-low crosslinked microgel (ULC), and $c=1\%$ and $c=5\%$ microgels. Corresponding b) convex-hull and c) surface mesh. d) Normalized distribution of the largest ellipsoid semi-axes obtained from the convex-hull and e) surface mesh representations for the three different systems at swollen conditions. Shape parameter, $S_p$, calculated from the f) convex hull and g) surface mesh.
}
\label{fig:sm_ch}
\end{figure*}

The method assumes isotropic elasticity, such that the strain energy can be expressed solely through the isotropic invariants of the deformation tensor. The $W$ functions from Eq.~\ref{eq:invariants} associated with each invariant can then be obtained from the potential of mean force (PMF) of their probability distributions as,
\begin{equation}
    W(I_x)=-k_\mathrm{B}T \ln\left[P(I_x)\right]+D_x,
    \label{eq:PMF}
\end{equation}
where $I_x$ denotes either $I_1$, $I_2$ or $J$, and $D_x$ is a constant. Then, the elastic moduli can be retrieved by fitting the PMF of each invariant with a function of the type $M_x(X-X_0)^\gamma+D_x$, where $M_x$, $X_0$ and $D_x$ are constants, $\gamma=2$ for $J$, since $J\approx 1$~\cite{Ninarello2023}, and $\gamma=1$ for $I_1=I_2=I$. 
The bulk and shear moduli are then given by

\begin{equation}
    K = \frac{2M_J}{V}, \,\, G = \frac{2M_I}{V},
\end{equation}

The remaining elastic moduli, namely, the Young's modulus ($Y$) and the Poisson's ratio ($\nu$), are calculated from $K$ and $G$ according to
\begin{equation}
    Y = \frac{9  K  G}{ 3  K + G},
\end{equation}

\begin{equation}
    \nu = \frac{3  K - 2  G} {2  (3  K + G)}.
\end{equation}

\noindent
The probability distributions of the strain invariants are first estimated by constructing histograms with the bin width determined according to the Freedman-Diaconis rule~\cite{freedman1981histogram}.
The sampled values of $I_1$ and $I_2$ are combined into one set $I$. Following Ref.~\cite{rovigatti2019connecting}, the fits of the PMFs are restricted to the region extending from the minimum of each potential up to $2k_\mathrm{B}T$ above it, where the strain-energy functions are well described by the fitting form.

\section{Results and Discussion}
\subsection{Shape fluctuations}

To uncover the elastic behavior of microgels, we examine three representative microgels that span a wide range of crosslinker concentrations, $c$: regular microgels with $c=5\%$ and $c=1\%$ and ULC microgels. 
We begin by characterizing the shape fluctuations of the three systems in the swollen state, $\alpha=0$, using two different representations: i) the convex hull enclosing all microgel monomers and ii) the surface mesh, previously employed for the structural characterization of ULCs~\cite{marin2026unexpected}. 
We show in Fig.~\ref{fig:sm_ch}a) representative configurations of each system. The ULC microgels are readily distinguished by the presence of long outer chains, a feature that is captured by both surface representations, shown in Fig.~\ref{fig:sm_ch}b) and c), respectively, and is expected to influence their elastic response.

\begin{figure*}[ht!]
\begin{center}
\includegraphics[width=1\linewidth]{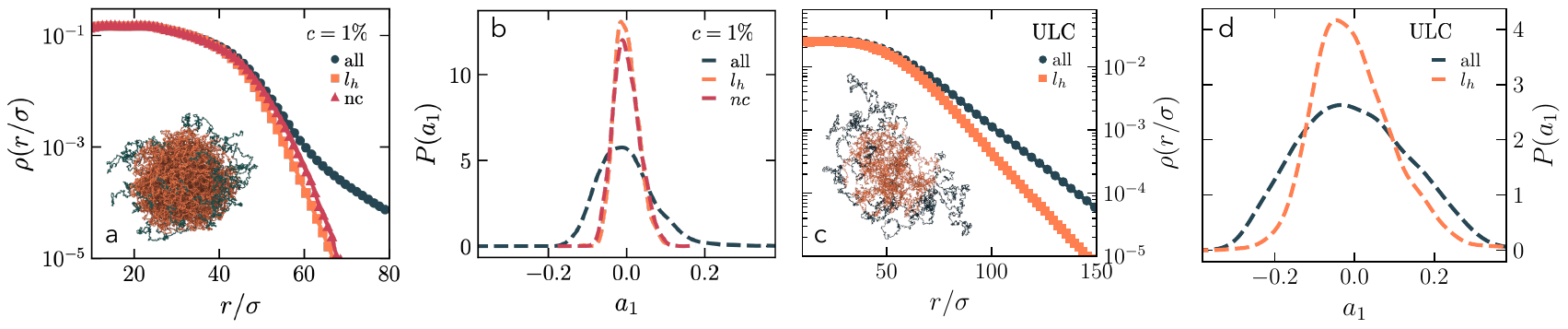}
\end{center}
\vspace{-0.5cm}
\caption{
\textbf{Effect of the outer chains} a) Radial density profile of a microgel of $c=1\%$ including all monomers (circles), accounting only for monomers within $r_{i,\mathrm{cm}}\leq l_h$ (squares), and for a microgel assembled without dangling ends (triangle symbols).  Inset shows a representative configuration of the $c=1\%$ full microgel, with monomers with $r_{i,\mathrm{cm}}\leq l_h$ colored in orange. b) Normalized probability distributions of the largest semi-axes of the ellipsoid associated with the gyration tensor for the three systems shown in a). c) $\rho(r)$ of the ULC with all monomers and without outer chains according to the $l_h$ criterion. The inset highlights the dangling chains. d) Corresponding distribution of the largest semi-axes. Removing the dangling chains restores an approximately Gaussian distribution. 
}
\label{fig:no_chain}
\end{figure*}

For regular microgels, both methods yield similar surface reconstructions due to their nearly spherical shape. In contrast, for ULCs, the convex hull encloses the long dangling chains and consequently overestimates the particle volume, whereas the surface mesh provides a tighter representation of the particle boundary. To build the latter, a probe radius must be defined depending on the topology of the system; we choose $R_\alpha=12\sigma$ for all systems, which provides a compromise between a detailed depiction of the outer chains and avoids the creation of spurious inner holes in its structure. 

After obtaining the shape with either the convex hull or the surface mesh, the corresponding gyration tensor is built and diagonalized to obtain its eigenvalues that can be related to the semi-axes $a_i$ of an ellipsoid, as $a_i=\sqrt{3\lambda_i}$, where $\lambda_i$ are the eigenvalues. To calculate the elastic moduli, the method based on Refs.~\cite{aggarwal2016nonuniform,rovigatti2019connecting} assumes that the particle fluctuates around a single equilibrium reference configuration with no strain. Indeed, if the system fluctuates isotopically around a reference configuration, the distribution in time of such semi-axes will have a normal distribution. To assess whether this assumption is satisfied, we analyze the distributions of the ellipsoid semi-axes obtained from the two surface representations. To enable comparison between all systems, we normalize the distributions so that they are centered at $0$. For the convex-hull representation, shown in Fig.~\ref{fig:sm_ch}d), both $c=1\%$ and $c=5\%$ exhibit approximately Gaussian distributions. Although ULC microgels display substantially broader fluctuations, their distribution remains unimodal, indicating that the assumption of fluctuations around a single equilibrium configuration is still satisfied.

The picture changes for the surface mesh, shown in Fig.~\ref{fig:sm_ch}e). While the $c=5\%$ microgel exhibits a narrower Gaussian distribution, both  $c=1\%$ and ULC cases develop clear bimodal distributions, allegedly coming from their outer chains. This indicates that the reconstructed surface fluctuates between two different conformations, thus violating the fundamental assumptions of the method.   

To further evaluate the validity of the isotropic approximation underlying both methods, we quantify the overall particle shape through the dimensionless shape parameter,
\begin{equation}
    S_p=\frac{V}{V_s}=6\sqrt{\pi}\frac{V}{A^{3/2}},
\end{equation}
where $V$ is the volume of the microgel obtained from its convex hull or surface mesh, and $V_s$ is the volume of a sphere with the same surface area as the microgel's convex hull or surface mesh. A value of $S_p=1$ corresponds to a perfect sphere, whereas smaller values indicate increasing deviations from sphericity. In Fig.~\ref{fig:sm_ch}f) and g), we show the $S_p$ of the convex hull and surface mesh for all three systems along the VPT, respectively. As expected, both methods yield $S_p>0.9$ for the $c=5\%$ microgel over the entire swelling curve, indicating an almost spherical particle. Differences between the two methods become increasingly pronounced as the crosslinker concentration decreases. In particular, the surface mesh shows a markedly nonspherical shape of ULCs in the swollen regime with $S_p<0.3$, reflecting how sensitive the method is to the extended outer chains. At $\alpha$ larger than the VPT temperature $\alpha\sim 0.63$, both methods gradually converge as the particle collapses. 

Overall, the surface mesh provides a more detailed description of the shape of the ULCs. However, the resulting shape fluctuations violate the assumptions required for our method used to calculate the elastic moduli, leading to noisy and bimodal distributions. Conversely, although the convex hull overestimates the volume of the ULCs, it preserves the fluctuations around a reference isotropic configuration. Therefore, it provides a more robust basis for the determination of the elastic moduli. Consequently, all elastic properties reported in the following are calculated with the convex hull method.

\subsection{Effect of outer chains}

One of the main reasons for the deviation from isotropic fluctuations may be due to the presence of long dangling chains, which are particularly pronounced in the ULC microgels. As well as affecting the validity of the fluctuation method, these chains may also contribute directly to the elastic response of the particle. To investigate their role, we first focus on the $c=1\%$ microgel, comparing its behavior before and after removing the dangling ends. 

We first establish a reference system in which the outer chains are removed during the assembly stage of a microgel with $c=1\%$. We directly remove the chains that are connected to the network through a single crosslinker, namely a loose end. After equilibrating the system, we compute the radial density profile, $\rho(r)$, of both the original microgel (`full') and the `no-outer-chain' (`nc') one. We show both $\rho(r)$ in Fig.~\ref{fig:no_chain}a); the dangling chains give rise to the low-density tail extending to large radial distances in the full microgel. 

Then, we test whether these chains can be identified directly from an equilibrated configuration using the structural criterion previously introduced in Refs.~\cite{del2024numerical,marin2026unexpected}. Specifically, particles that are at a distance from the center of mass of $r_{i,\mathrm{cm}}>l_h$ are identified as part of the outer chains, where $l_h=R_c+\sigma_s$. Here, $R_c$ and $\sigma_s$ correspond to the core radius and the half-width of the corona, respectively, obtained by fitting the density profile accounting for all monomers with the `so-called' fuzzy-sphere model. 

\begin{figure*}[ht!]
\begin{center}
\includegraphics[width=1\linewidth]{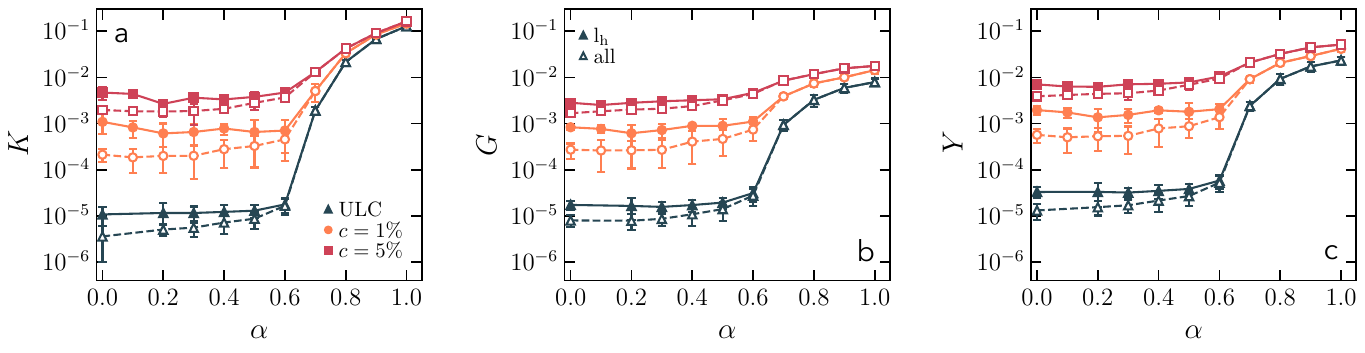}
\end{center}
\vspace{-0.5cm}
\caption{
\textbf{Elastic Moduli} a) Bulk b) Shear and c) Young's moduli 
as a function of the parameter $\alpha$, corresponding to an effective temperature, of ultra-low crosslinked (ULC) microgels (triangles) and  $c=1\%$ (circles) and $c=5\%$ (squares) microgels. Lines are guides for the eye. Open symbols with dashed lines correspond to the calculation accounting for all monomers, and closed symbols with continuous lines to the systems without outer chains. 
}
\label{fig:moduli}
\end{figure*}

The monomers identified as outer chains with $l_h$ as the cutoff are depicted in the inset of Fig.~\ref{fig:no_chain}a) in green, while the others are orange. To validate this approach, we recalculate the density profile accounting for only the monomers with $r_{i,\mathrm{cm}}\leq l_h$ and compare with that of the `no-outer-chain' case. As shown in Fig.~\ref{fig:no_chain}a), both profiles are in close agreement, demonstrating that the cutoff $l_h$ can be used as a good criterion for the identification of such chains.

The next step is to examine how the shape fluctuations are affected by the removal of the dangling chains once they have been identified. To do so, we compute the convex hull of the three cases and the semi-axes of the ellipsoid related to the corresponding gyration tensor. We show in Fig.~\ref{fig:no_chain}b) the probability distributions of the largest semi-axis $a_1$ for each. Removing the dangling chains, either explicitly during assembly or through the $l_h$ criterion, dramatically narrows the distribution, and in both cases the distributions recover an approximately Gaussian shape, indicating that those broad fluctuations observed in the full microgel originate predominantly from the dangling chains rather than from the network backbone.

In the case of ULCs, although these systems lack a well-defined core, their density profiles are still accurately described by the fuzzy-sphere model, allowing $l_h$ to be determined in the same manner. We show in Fig.~\ref{fig:no_chain}c) the ULC density profile accounting for all monomers and of the system after removing the outer chains, respectively. As expected, the dangling chains extend considerably farther from the particle center than in the $c=1\%$ microgel, but removal substantially modifies the semi-axes distribution. Although the fluctuations remain broader than those of the $c=1\%$, the distributions recover an approximately Gaussian shape, as shown in Fig.~\ref{fig:no_chain}d). These results demonstrate that the large, non-Gaussian shape fluctuations characteristic of low-crosslinked microgels arise predominantly from the highly mobile dangling chains.

\subsection{Elastic moduli}

Having established that the convex-hull representation provides a reliable description of the shape fluctuations, we now investigate how the elastic response evolves across the volume phase transition for microgels with different crosslinker concentrations. We compare the elastic moduli calculated using all monomers with those obtained after removing the dangling chains. We show in Fig.~\ref{fig:moduli} the bulk ($K$), shear ($G$), and Young's moduli ($Y$), 
respectively. The first thing to notice is that by removing the outer chains, the three moduli systematically increase, indicating that these highly mobile regions soften the overall mechanical response. However, the effect is relatively small and does not alter the evolution of the elastic response across the VPT. As expected, once the microgels collapse, above the VPT temperature, both approaches converge, yielding nearly identical elastic moduli. Note that $l_h$ is defined from the swollen-state structure of each microgel and kept fixed throughout the analysis. 

The ULC exhibits a markedly `softer' nature as compared to the other two microgels. In the swollen state, $\alpha=0$, the elastic moduli are almost two orders of magnitude smaller than those of $c=5\%$ and remain substantially lower than those of $c=1\%$.
However, above the VPT temperature, the elastic response becomes nearly independent of the crosslinker concentration. Previous studies reported a monotonic increase of the bulk modulus with both $\alpha$ and $c$~\cite{rovigatti2019connecting}. However, we observe that the behavior of the moduli upon crossing the VPT becomes increasingly steep as the crosslinker concentration decreases, being particularly sharp for the ULC microgels. However, in the present data, we do not observe the presence of a minimum either in $K$ or in $G$ close to the VPT temperature,  differently from earlier experimental results~\cite{voudouris2013micromechanics,hashmi2009mechanical,burmistrova2011effect}. This could be due to the small size of the simulated microgels as compared to the quasi-millimetric size of those studied by capillary micromechanics~\cite{voudouris2013micromechanics}, thus more similar to hydrogels for which a minimum in $K$ was indeed observed~\cite{Ninarello2025}. In addition, AFM measurements~\cite{hashmi2009mechanical,burmistrova2011effect} of individual microgels are notoriously difficult~\cite{scotti2022softness,schulte2022microgels}, 
so that additional investigations are needed to clarify these points.

\subsection{Signatures of auxetic behavior}

\begin{figure}[ht!]
\begin{center}
\includegraphics[width=0.8\linewidth]{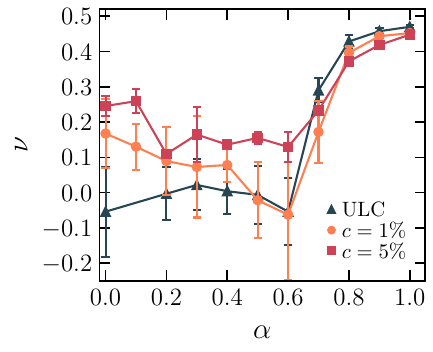}
\end{center}
\vspace{-0.5cm}
\caption{
\textbf{Poisson's ratio} Poisson's ratio accounting for the systems without outer chains,
as a function of the parameter $\alpha$ of ultra-low crosslinked (ULC) microgels (triangles) and  $c=1\%$ (circles) and $c=5\%$ (squares) microgels. Lines are guides for the eye. 
}
\label{fig:poisson's}
\end{figure}

We finally turn to examine the behavior of the Poisson's ratio to detect the possibility of auxetic behavior in microgels.
The evolution of $\nu$ with temperature is reported in Fig.~\ref{fig:poisson's}.
Starting with regular microgels, we find that both $c=5\%$ and
$c=1\%$ cases display a positive Poisson's ratio in good solvent, developing a minimum close to the VPT. In addition, the value of $\nu$ decreases with $c$, in agreement with earlier studies on smaller microgels\cite{rovigatti2019connecting}. Interestingly, we find that 
for $c=1\%$ the Poisson's ratio becomes slightly negative close to the VPT. This indicates that auxeticity emerges progressively as the network connectivity decreases. 
Interestingly, this trend is qualitatively consistent with the auxetic response reported for weakly crosslinked hydrogels under tensile loading~\cite{Ninarello2022}. Although the present analysis is based on equilibrium shape fluctuations rather than externally applied deformation, both systems suggest that reducing the network connectivity promotes auxetic behavior.

Turning to the ULC microgels, we strikingly find that $\nu <0$ at all temperatures below and up to the VPT. This behavior reflects the markedly different behavior of the bulk and shear moduli with respect to the regular microgels. Indeed, for ULCs, the bulk modulus becomes comparable to, and in some cases even smaller than, the shear modulus throughout the swollen regime and near the VPT. Within isotropic elasticity, this leads to a sufficiently small $K/G$ ratio causing the Poisson's ratio, $\nu$, approach to zero. Eventually, it even becomes negative once $K<2G/3$, providing the mechanical signature of auxeticity.

Since the Poisson's ratio is derived from both $K$ and $G$, it is particularly sensitive to statistical fluctuations. This effect is especially pronounced for ULC microgels because of their intrinsically large shape fluctuations. We remark that in order to improve statistical accuracy, all ULC results were averaged over ten independently assembled networks. 

Finally, upon complete collapse, all systems converge towards $\nu\approx0.4$, close to the value reported for crystalline hard-sphere solids~\cite{tretiakov2005poisson}. This behavior is consistent with the dense, sphere-like character of collapsed microgels. Overall, these results demonstrate that reducing the crosslinker concentration progressively lowers the ratio $K/G$, driving the Poisson's ratio from positive values toward zero and eventually into an auxetic regime.

\subsection{Auxeticity and inner structure}

Finally, we investigate how the elastic response varies inside the microgel.
To this end, we probe the local elastic response by progressively restricting the fluctuation analysis to monomers located within a progressively smaller distance,
\begin{equation}
     r_{i,\mathrm{cm}}\leq R_c+n\sigma_s/10,
\end{equation}
where $n$ goes from $n\in[-1,2]$, and $R_c$ and $\sigma_s$ are obtained from the fuzzy-sphere model fit of the density profiles.
In this way, we successively include larger fractions of the particle, ranging from the inner core to the complete microgel. We remark that for ULC microgels, no real core is detected~\cite{marin2026unexpected}. 

\begin{figure}[h!]
\begin{center}
\includegraphics[width=0.8\linewidth]{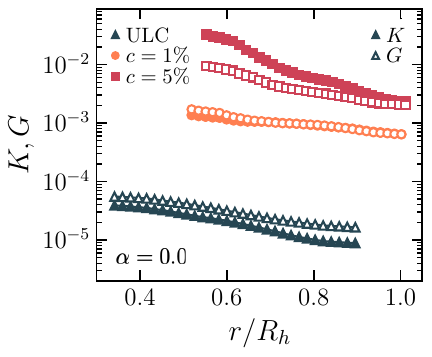}
\end{center}
\vspace{-0.5cm}
\caption{
\textbf{Elastic response within the microgel} Bulk modulus, $K$, (filled symbols) and shear modulus, $G$, (empty symbols) as functions of the radial distance $r$ to the core of the particle normalized by the corresponding hydrodynamic radius, $R_h$, for ULC (triangles), $c=1\%$ (circles) and $c=5\%$ (squares) microgels. All systems are in the swollen state, $\alpha=0$.
}
\label{fig:rdepence}
\end{figure}

We show in Fig.~\ref{fig:rdepence} the dependence of $K$ and $G$ as a function of $r/R_h$, where $R_h$ is the microgel hydrodynamic radius (see Methods), for the three studied microgels in the swollen state with $\alpha=0.0$. For all systems, both moduli increase as the analysis is restricted toward the particle center, indicating that the core is mechanically stiffer than the outer corona. In particular, this effect is most pronounced for the $c=5\%$ microgel, whose dense core produces an increase of almost one order of magnitude in the bulk modulus, in agreement with estimates based on neutron scattering experiments of microgels~\cite{houston2022resolving}. Throughout the particle, the bulk modulus remains larger than the shear modulus, and the difference between the two becomes increasingly pronounced toward the particle center.
By contrast, the radial dependence becomes progressively weaker as the crosslinker concentration decreases. In particular, ULCs only exhibit a modest increase in both moduli with decreasing $r$, demonstrating that their elastic response is roughly homogeneous throughout the particle. For $c=1$\% microgels, the situation is intermediate between the other two cases: it is more similar to the ULC case than to $5\%$ microgels, with a mild increase of the moduli within the microgel, although the value of the moduli is much closer to that of the more crosslinked microgels. Interestingly, in the case of the ULC, the shear modulus remains larger than the bulk modulus over the entire particle, consistent with a negative Poisson's ratio.

\begin{figure}[t!]
\begin{center}
\includegraphics[width=0.8\linewidth]{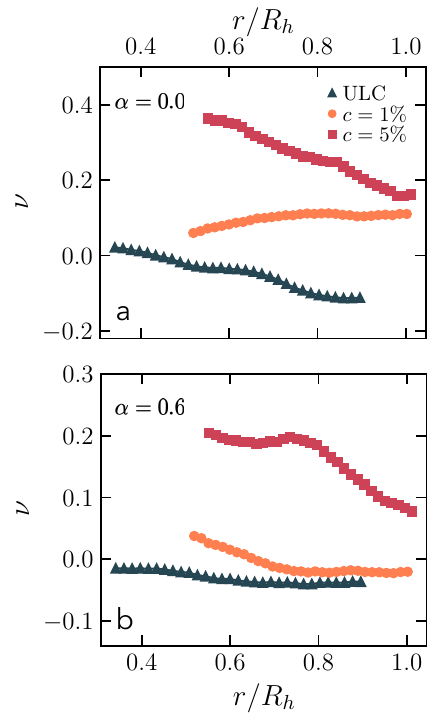}
\end{center}
\vspace{-0.5cm}
\caption{
\textbf{Radial dependence of Poisson ratio's} Poisson's ratio, $\nu$, as a function of the normalized distance $r/R_h$ for $c=5\%$ (squares), $c=1\%$ (circles), and ULC (triangles) microgels at a) $\alpha=0.0$ and b) $\alpha=0.6$, corresponding to the state point below the VPT where the minimum in $\nu$ is observed. 
}
\label{fig:poisson_ulc}
\end{figure}

Next, we analyze how the auxetic response is distributed within the particle. To address this, we calculate the Poisson's ratio at the swollen state, $\alpha=0.0$, and at $\alpha=0.6$, close to the VPT, where the strongest auxetic response is observed. 
We show $\nu$ as a function of $r/R_h$ in Fig.~\ref{fig:poisson_ulc}. For both temperatures, the $c=5\%$ microgel exhibits a positive $\nu$ throughout the particle, with the largest values corresponding to the core of the particle, as expected. By contrast, the microgel with $c=1\%$ displays a positive Poisson's ratio in the swollen state. Here, a slight decrease in $\nu$ is observed close to the center of the particle, possibly due to statistical noise. On the other hand, for $\alpha=0.6$, it develops a narrow region with a negative $\nu$, corresponding to the outer corona, while the core remains weakly positive with $\nu$ reaching a value $\sim 0.1$.

A qualitatively different behavior is observed for the ULC microgels. For both temperatures, the Poisson's ratio remains negative over the entire particle range, demonstrating that the auxetic response is not confined to a particular structural region but is instead an intrinsic property of the weakly connected polymer network. This is in line with the homogeneous, loose structure of ULCs, which do not possess a marked core, thus being able to respond auxetically independently of the local environment.

\section{Conclusions}
In this work, we have investigated the elastic response of microgels with different crosslinker concentrations across the volume phase transition, ranging from regular microgels with $c=5\%$ to low-crosslinked ones with $c=1\%$ and also ultra-low-crosslinked microgels ($c=0.1\%$ from self-crosslinking among monomers~\cite{gao2003cross}). To determine the elastic moduli, we employed a fluctuation-based approach based on the equilibrium shape fluctuations of individual microgels. We assessed two different representations of the microgel volume, namely the convex hull and the surface mesh. Although the surface mesh provides a more faithful description of the loose morphology of ULC microgels, it violates the central assumption of the method, namely that the particle fluctuates around a well-defined reference configuration. By contrast, the convex-hull representation preserves approximately Gaussian fluctuations of the ellipsoidal semi-axes and therefore provides a robust framework for determining the elastic moduli.
We also demonstrated that the external dangling chains, which are extremely common in microgels with low $c$, further broaden the shape fluctuations without significantly modifying the elastic response itself. Therefore, their removal improves the robustness of the fluctuation analysis while preserving the elastic behavior across the volume phase transition.

Having established the applicability of the fluctuation method to weakly connected microgels, we then evaluated the elastic properties and, for the first time, reported the onset of auxetic behavior in microgels when the network connectivity decreases. Indeed, for $c=1\%$ the Poisson's ratio develops a negative minimum at the VPT. In addition, for ULC microgels, the auxetic behavior extends at all temperatures below and up to the VPT.
By further resolving the elastic response within different regions of the particle, we demonstrate that the auxetic response is not confined to the outer corona or to the dangling chains. Instead, it extends throughout the weakly connected polymer network, indicating that auxeticity is an intrinsic property of the internal architecture rather than a consequence of surface fluctuations alone.

The present findings also have important implications for coarse-grained descriptions of soft colloids. Many theoretical models of soft particles, including microgels, assume a fixed value of the Poisson's ratio~\cite{Boon2017,alade2026modeling}. Our results demonstrate that $\nu$ strongly depends on both the crosslinker concentration and the thermodynamic state of the particle, approaching hard-sphere-like values only in the fully collapsed regime. Consequently, accounting for this state dependence may be important for accurately describing the collective behavior of soft colloids.

Finally, the present results complement earlier works on smaller microgels at higher $c$~\cite{rovigatti2019connecting}, for which a minimum of $\nu$ at the VPT was observed for $c=3.2\%$ but still of positive sign. In addition, they extend the observations of auxeticity in hydrogels, but only occurring under the application of a small tension~\cite{Ninarello2022}. Here, instead, the negative Poisson's ratio is found to be present at zero external pressure, under equilibrium conditions. We hope that these findings will stimulate new experimental efforts towards measuring elastic properties of individual microgels at very low crosslinking, particularly ULC ones, not yet available in the literature.

\section*{Conflicts of interest}
There are no conflicts to declare.

\section*{Data availability}
Supporting data for this article will be made available at Zenodo.

\section*{Acknowledgements}
We thank Lorenzo Rovigatti and Andrea Ninarello for useful discussions.
SMA and EZ acknowledge funding by the European Union HORIZON-MSCA-2022-Postdoctoral Fellowships under grant agreement no. 101106848, MGELS. 
LR and EZ acknowledge financial support from the European Union (HorizonMSCA-Doctoral Networks) through the project QLUSTER (HORIZON-MSCA-2021-DN-01-GA101072964). We gratefully acknowledge the CINECA award under the ISCRA initiative for the availability of high-performance computing resources and support.  




\bibliography{elas} 
\bibliographystyle{rsc} 

\end{document}